\documentclass[aps,prl,preprint,showpacs,groupedaddress,showpacs]{revtex4}

\usepackage{graphicx}
\usepackage{dcolumn}
\usepackage{amssymb}
\usepackage{bm}

\newcommand{\wprime}{\mbox{$W^{\prime}$}}
\newcommand{\zprime}{\mbox{$Z^{\prime}$}}

\newcommand{\qbar}{\mbox{$\overline{q}$}}

\newcommand{\pbarp}{\mbox{$\overline{p}p$}}
\newcommand{\qstar}{\mbox{$q^{\ast}$}}
\newcommand{\ipb}{\mbox{pb$^{-1}$}}
\newcommand{\gevcc}{\mbox{GeV/$c^2$}}

\begin{document}

\title{Search for New Particles in the Two-Jet Decay Channel with
the D\O\ Detector}

\date{13 August 2003}
%
\author{                                                                      
V.M.~Abazov,$^{21}$                                                           
B.~Abbott,$^{55}$                                                             
A.~Abdesselam,$^{11}$                                                         
M.~Abolins,$^{48}$                                                            
V.~Abramov,$^{24}$                                                            
B.S.~Acharya,$^{17}$                                                          
D.L.~Adams,$^{53}$                                                            
M.~Adams,$^{35}$                                                              
S.N.~Ahmed,$^{20}$                                                            
G.D.~Alexeev,$^{21}$
% ...
\\                                                                            
\vskip 0.30cm                                                                 
\centerline{(D\O\ Collaboration)}                                             
\vskip 0.30cm                                                                 
}                                                                             
\address{                                                                     
\centerline{$^{1}$Universidad de Buenos Aires, Buenos Aires, Argentina}       
\centerline{$^{2}$LAFEX, Centro Brasileiro de Pesquisas F{\'\i}sicas,         
                  Rio de Janeiro, Brazil}                                     
\centerline{$^{3}$Universidade do Estado do Rio de Janeiro,                   
                  Rio de Janeiro, Brazil}                                     
\centerline{$^{4}$Institute of High Energy Physics, Beijing,                  
                  People's Republic of China}                                 
\centerline{$^{5}$Universidad de los Andes, Bogot\'{a}, Colombia}             
\centerline{$^{6}$Charles University, Center for Particle Physics,            
                  Prague, Czech Republic}                                     
\centerline{$^{7}$Institute of Physics, Academy of Sciences, Center           
                  for Particle Physics, Prague, Czech Republic}               
\centerline{$^{8}$Universidad San Francisco de Quito, Quito, Ecuador}         
\centerline{$^{9}$Laboratoire de Physique Subatomique et de Cosmologie,       
                  IN2P3-CNRS, Universite de Grenoble 1, Grenoble, France}     
\centerline{$^{10}$CPPM, IN2P3-CNRS, Universit\'e de la M\'editerran\'ee,     
                  Marseille, France}                                          
\centerline{$^{11}$Laboratoire de l'Acc\'el\'erateur Lin\'eaire,              
                  IN2P3-CNRS, Orsay, France}                                  
\centerline{$^{12}$LPNHE, Universit\'es Paris VI and VII, IN2P3-CNRS,         
                  Paris, France}                                              
\centerline{$^{13}$DAPNIA/Service de Physique des Particules, CEA, Saclay,    
                  France}                                                     
\centerline{$^{14}$Universit{\"a}t Mainz, Institut f{\"u}r Physik,            
                  Mainz, Germany}                                             
\centerline{$^{15}$Panjab University, Chandigarh, India}                      
\centerline{$^{16}$Delhi University, Delhi, India}                            
\centerline{$^{17}$Tata Institute of Fundamental Research, Mumbai, India}     
\centerline{$^{18}$CINVESTAV, Mexico City, Mexico}                            
\centerline{$^{19}$FOM-Institute NIKHEF and University of                     
                  Amsterdam/NIKHEF, Amsterdam, The Netherlands}               
\centerline{$^{20}$University of Nijmegen/NIKHEF, Nijmegen, The               
                  Netherlands}                                                
\centerline{$^{21}$Joint Institute for Nuclear Research, Dubna, Russia}       
\centerline{$^{22}$Institute for Theoretical and Experimental Physics,        
                   Moscow, Russia}                                            
\centerline{$^{23}$Moscow State University, Moscow, Russia}                   
\centerline{$^{24}$Institute for High Energy Physics, Protvino, Russia}       
\centerline{$^{25}$Lancaster University, Lancaster, United Kingdom}           
\centerline{$^{26}$Imperial College, London, United Kingdom}                  
\centerline{$^{27}$University of Arizona, Tucson, Arizona 85721}              
\centerline{$^{28}$Lawrence Berkeley National Laboratory and University of    
                  California, Berkeley, California 94720}                     
\centerline{$^{29}$University of California, Davis, California 95616}         
\centerline{$^{30}$California State University, Fresno, California 93740}     
\centerline{$^{31}$University of California, Irvine, California 92697}        
\centerline{$^{32}$University of California, Riverside, California 92521}     
\centerline{$^{33}$Florida State University, Tallahassee, Florida 32306}      
\centerline{$^{34}$Fermi National Accelerator Laboratory, Batavia,            
                   Illinois 60510}                                            
\centerline{$^{35}$University of Illinois at Chicago, Chicago,                
                   Illinois 60607}                                            
\centerline{$^{36}$Northern Illinois University, DeKalb, Illinois 60115}      
\centerline{$^{37}$Northwestern University, Evanston, Illinois 60208}         
\centerline{$^{38}$Indiana University, Bloomington, Indiana 47405}            
\centerline{$^{39}$University of Notre Dame, Notre Dame, Indiana 46556}       
\centerline{$^{40}$Iowa State University, Ames, Iowa 50011}                   
\centerline{$^{41}$University of Kansas, Lawrence, Kansas 66045}              
\centerline{$^{42}$Kansas State University, Manhattan, Kansas 66506}          
\centerline{$^{43}$Louisiana Tech University, Ruston, Louisiana 71272}        
\centerline{$^{44}$University of Maryland, College Park, Maryland 20742}      
\centerline{$^{45}$Boston University, Boston, Massachusetts 02215}            
\centerline{$^{46}$Northeastern University, Boston, Massachusetts 02115}      
\centerline{$^{47}$University of Michigan, Ann Arbor, Michigan 48109}         
\centerline{$^{48}$Michigan State University, East Lansing, Michigan 48824}   
\centerline{$^{49}$University of Nebraska, Lincoln, Nebraska 68588}           
\centerline{$^{50}$Columbia University, New York, New York 10027}             
\centerline{$^{51}$University of Rochester, Rochester, New York 14627}        
\centerline{$^{52}$State University of New York, Stony Brook,                 
                   New York 11794}                                            
\centerline{$^{53}$Brookhaven National Laboratory, Upton, New York 11973}     
\centerline{$^{54}$Langston University, Langston, Oklahoma 73050}             
\centerline{$^{55}$University of Oklahoma, Norman, Oklahoma 73019}            
\centerline{$^{56}$Brown University, Providence, Rhode Island 02912}          
\centerline{$^{57}$University of Texas, Arlington, Texas 76019}               
\centerline{$^{58}$Rice University, Houston, Texas 77005}                     
\centerline{$^{59}$University of Virginia, Charlottesville, Virginia 22901}   
\centerline{$^{60}$University of Washington, Seattle, Washington 98195}       
}                                                                             

\begin{abstract}
 We present the results of a search for the production of new
 particles decaying into two jets in \pbarp\ collisions at $\sqrt{s}$
 = 1.8 TeV, using the D\O\ 1992--1995 data set corresponding to 109
 \ipb . We exclude at the 95\% confidence level the production of
 excited quarks (\qstar ) with masses below 775 \gevcc , the most
 restictive limit to date. We also exclude standard-model-like
 \wprime\ (\zprime ) bosons with masses between 300 and 800 \gevcc\
 (400 and 640 \gevcc ). A \wprime\ boson with mass $<$ 300 \gevcc\ has
 been excluded by previous measurements, and our lower limit is
 therefore the most stringent to date.
\end{abstract}

%
%
\maketitle

 The direct production of hadronic jets is the dominant contribution
 to high transverse momentum ($p_{T}$) processes in antiproton--proton
 (\pbarp ) collisions. There are many extensions of the standard model
 that predict the existence of new massive objects ({\rm e.g.},
 excited quarks, \wprime\ and \zprime\
 bosons~\cite{excited_quarks,ext_gauge}) that couple to quarks and/or
 gluons and may be observed as resonant structures in the two-jet mass
 spectrum.  The previous observation of $W$ and $Z$ bosons decaying
 into two jets in the UA2 experiment~\cite{UA2wjj} proved the
 feasibility of doing dijet mass spectroscopy at $\overline{p}p$
 colliders. Subsequently, the UA2~\cite{ua2xjj} and CDF~\cite{cdfxjj}
 experiments searched for new resonances in the dijet mass spectrum,
 and set limits on their production within the context of different
 theoretical models.  This paper reports on a search for such
 resonances in the two-jet mass spectrum~\cite{d0_dijet,d0_dijet2} using the
 data collected at a center-of-mass energy of 1.8~TeV with the D\O\
 detector in 1992--1995, corresponding to an integrated luminosity of
 109 \ipb .

 Jet detection in the D\O\ detector~\cite{d0} primarily utilizes the
 uranium/liquid--argon calorimeters that cover the pseudorapidity
 region $\mid \!  \eta \! \mid \lesssim 4$, where $\eta =
 -\ln[\tan(\theta/2)]$ and $\theta$ is the polar angle with respect to
 the proton beam. Jets are reconstructed using an iterative jet cone
 algorithm with a cone radius of $\cal{R}$=0.7 in $\eta$--$\phi$
 space~\cite{d0_dijet}, where $\phi$ is the azimuthal
 angle. Background jets from isolated noisy calorimeter cells and
 accelerator losses are minimized via jet-quality
 criteria ~\cite{d0_dijet}.  The transverse energy of each jet is then
 corrected~\cite{energy_scale} for offsets due to the underlying
 event, noise, multiple interactions and pileup, the fraction of
 particle energy showering outside of the jet cone, and calorimeter
 energy response to incident hadrons.

 For each event that passes the quality criteria, the inclusive dijet
 mass can be calculated, assuming that the particles within the jets
 are massless, using the relationship $M^{2} = 2 E_{T}^{1} E_{T}^{2} [
 \cosh( \Delta \eta ) - \cos ( \Delta \phi )]$, where $E_{T}^1$ and
 $E_{T}^2$ are the transverse energies of the two highest-$E_T$
 jets. Each event is then weighted by the inverse of the efficiency of
 the quality criteria. The pseudorapidities of the two leading jets
 are selected to be $\mid \!\eta_{1,2} \!  \mid < 1.0$ and $\Delta
 \eta = \mid \! \eta_{1} - \eta_{2} \! \mid < 1.6$ in order to
 maximize the range of dijet mass at which the trigger is efficient.

\begin{figure}[htbp]
\includegraphics[width=\columnwidth ]{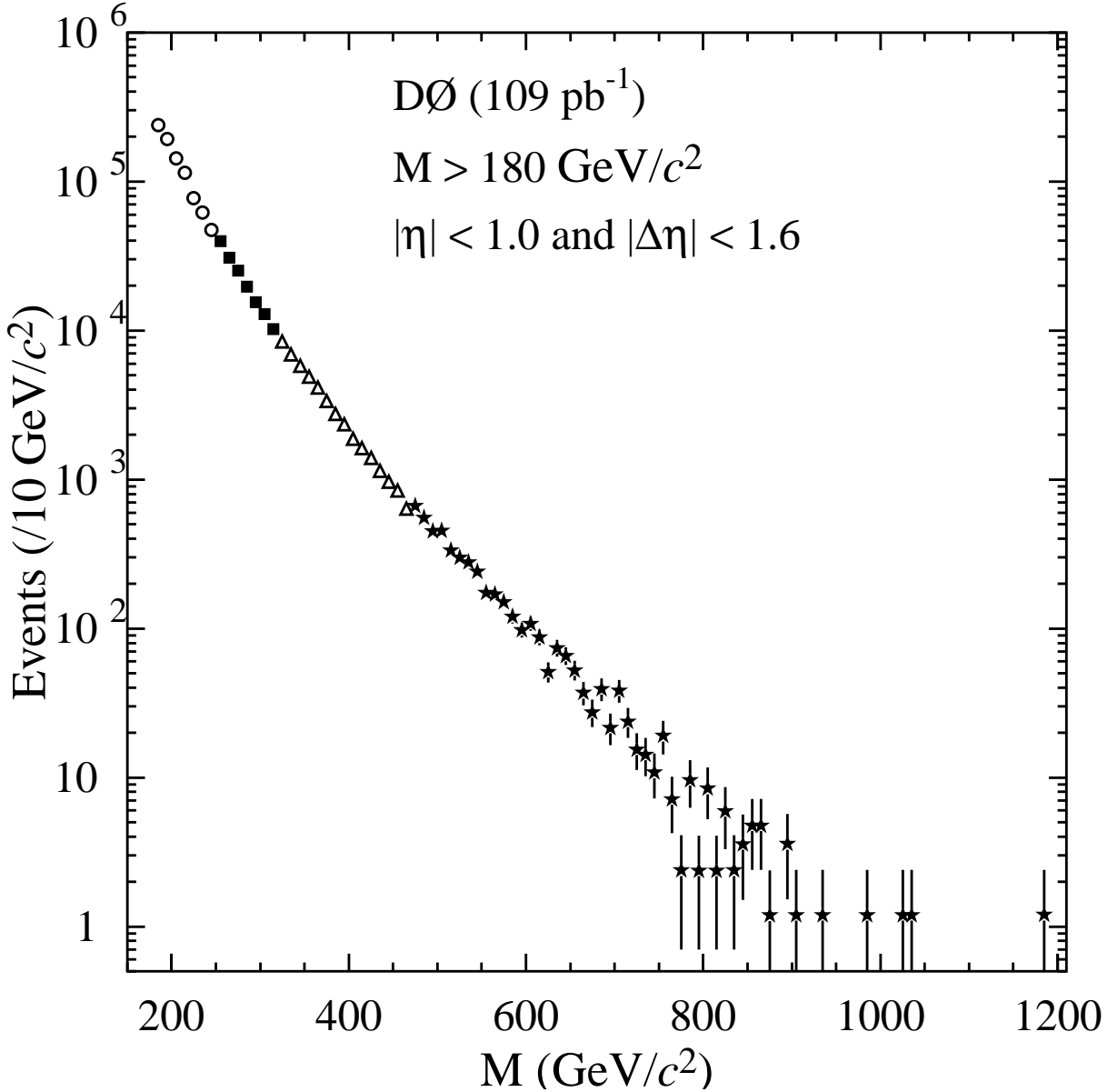}
\caption{\label{dijet_mass} The inclusive dijet mass spectrum. The
 events from each trigger have been corrected by the trigger's
 luminosity and event efficiency. The data were collected using
 triggers with uncorrected $E_T$ thresholds of 30 (open circles), 50
 (solid squares), 85 (open triangles), and 115 GeV (solid stars). The
 error bars represent statistical uncertainties.}
\end{figure}

 A single trigger was used to collect the 1992--1993 data, with an
 $E_T$ threshold of 115 GeV, for an integrated luminosity of 14.1 \ipb
 .  During 1994--1995, the data were collected using four triggers,
 with uncorrected $E_T$ thresholds of 30, 50, 85 and 115 GeV, for
 integrated luminosities of 0.36, 4.8, 56.5, and 94.9 \ipb ,
 respectively.  After the jet-energy corrections, these trigger
 samples are used to measure the dijet mass spectrum above mass
 thresholds of 180, 250, 320, and 470 GeV, respectively, where each of
 the triggers is $> 97\%$ efficient. The resulting dijet mass spectrum
 is shown in Fig.~\ref{dijet_mass}. The widths of the mass bins are
 chosen such that all events in any bin are recorded using a single
 trigger, there were $> 10$ events per bin, and the bin width is
 approximately equal to the mass resolution.

 The uncertainty in the dijet mass spectrum from the uncertainty in
 luminosity is $5.8\%$, and the uncertainty from the jet-quality and
 vertex criteria is $1\%$. The uncertainties due to the jet energy
 scale~\cite{energy_scale} are 7$\%$ (30$\%$) for the lowest (highest)
 mass bin, and are correlated.  The uncertainty in energy scale has
 three components: the uncorrelated, fully correlated, and
 partially-correlated uncertainties.  A correlation matrix is
 calculated for the partially correlated uncertainties using the
 method described in Ref.~\cite{d0_dijet}. The uncertainties in the
 mass spectra due to the jet energy resolution are (0.5--3.0)$\%$ over
 the mass range under consideration.

 We consider three models for a possible signal in the dijet mass
 spectrum. The first model contains a mass-degenerate family of
 excited quarks~\cite{excited_quarks} that decay to a quark and a
 gluon ($\qstar \rightarrow qg$).  We assume that the coupling
 parameters of the excited quarks equal unity ($f = f^{\prime} = f_{s}
 = 1$) and that the compositeness scale equals the mass of the excited
 quark ($\Lambda^{\ast} = M_{q^{\ast}}$). The second and third
 models~\cite{ext_gauge} contain additional $W$ and $Z$ bosons,
 respectively, with standard-model-like couplings, where all possible
 quark decays are allowed ($\wprime \rightarrow q\qbar^{\prime}$,
 $\zprime \rightarrow q\qbar$, with $\wprime \rightarrow t\bar{b}$,
 and $\zprime \rightarrow t\bar{t}$ allowed when kinematically
 possible). The leading-order \wprime\ and \zprime\ boson production
 cross sections are corrected by NLO ``$K$ factors''~\cite{eppley} of
 approximately 1.3, to account for higher-order effects.  All models
 were generated using {\sc pythia} 6.2 ~\cite{pythia}, with the {\sc
 cteq6} parton distribution functions (PDFs) \cite{cteq6}.

\begin{figure}[htbp]
\includegraphics[width=\columnwidth ]{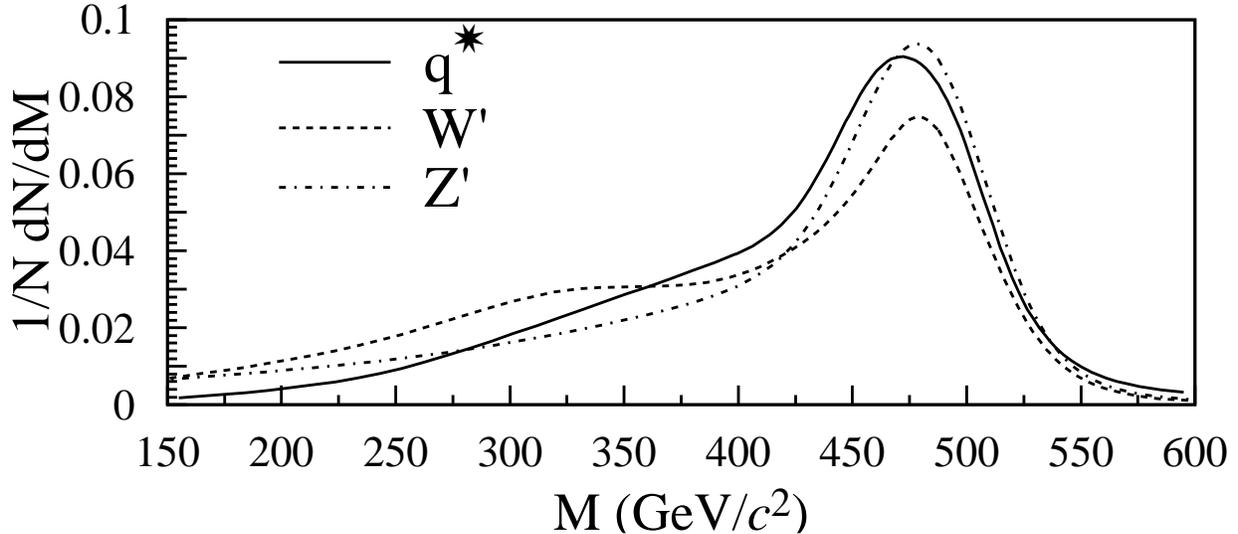}
\caption{ \label{line_shape_fig} The line shapes of a 500 GeV/$c^2$
 \qstar , \wprime\ and \zprime\ bosons , smoothed and normalized to
 unit area.}
\end{figure}

 For each of the models, a Monte Carlo mass spectrum was generated at
 25 GeV/$c^2$ intervals from a mass of 150 GeV/$c^2$ to 1
 TeV/$c^2$. Jets are reconstructed at the particle level using the
 same iterative jet cone algorithm that is applied to the data. The
 resulting energies are then smeared with the measured jet
 resolutions. Each of the mass spectra contains 50,000
 events. Examples of the spectra generated for a resonant mass of 500
 GeV/$c^2$ are shown in Fig.~\ref{line_shape_fig}.

 The data were analyzed using a Bayesian technique, with a flat prior
 for the signal (see Ref.~\cite{d0conlimits}). The predicted number of
 events per bin is given by $\mu_i =
 \left(\sigma_{{QCD}_{i}}{C_{QCD_{i}}} + N_{X_i} \sigma_{X}{C_{X_{i}}}
 \right) {\cal{L}}_{i} {\epsilon_{i}} $ where $\sigma_{{QCD}_{i}}$ is
 the predicted QCD two-jet cross section for mass bin $i$; $N_{X_i}$
 is the fraction of signal events in the bin ($\sum N_{X_i} = 1$);
 $\sigma_X$ is the cross section for the signal; $\mathcal{L}_i$ is
 the integrated luminosity; $\epsilon_{i}$ corresponds to the product
 of the efficiencies of the jet-quality criteria, the vertex selection
 efficiencies, and the trigger efficiencies per bin; and
 ${C_{QCD_{i}}}$ and ${C_{X_{i}}}$ are the jet energy and resolution
 corrections on the background and signal, respectively. Assuming
 $N_i$ follows Poisson statistics, the probability that $N_i$ events
 are observed in a given mass bin is then given by $P \left( N_i \mid
 \sigma_{{QCD}_{i}}, \sigma_{X}, N_{X_i}, {\cal L},
 \epsilon_{i},{C_{X_{i}}},{C_{QCD_{i}}},I \right) = { {e^{-\mu_i}
 {\mu_i}^{N_i}} / {N_{i}!}}$, where $I$ reflects all other
 ``nuisance'' parameters. The probability of observing the set $N_i$
 that makes up the mass spectrum is then given by the product of these
 probabilities. To calculate the probability distribution for
 $\sigma_X$, Bayes' theorem is applied with the following assumptions
 about the prior probability distributions: $\sigma_X$ has a uniform
 prior; $\sigma_{{QCD}_{i}}$, $\epsilon_{i}$, ${C_{QCD_{i}}}$,
 ${C_{X_{i}}}$ and ${\cal{L}}_{i}$ all have Gaussian priors with
 widths given by their uncertainties; and $N_{X_i}$ has a Poisson
 prior.

\begin{figure}[htbp]
\includegraphics[width=\columnwidth ]{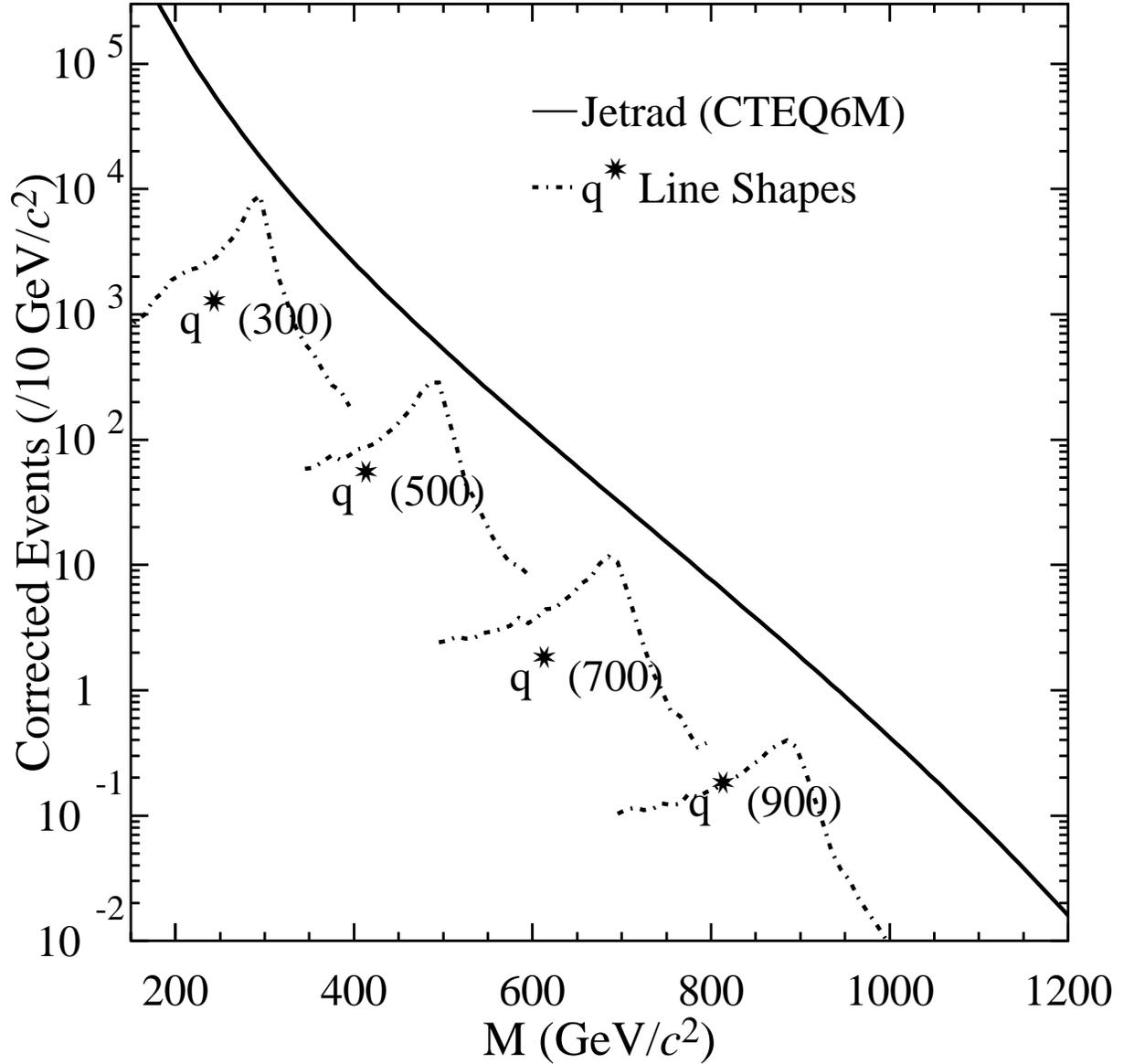}
\caption{\label{jetrad_fig} The {\sc Jetrad} (solid line) simulation
 of the inclusive dijet mass spectrum. The dashed-dotted lines show
 {\sc Pythia} simulations of the excited quark line shapes for
 $M_{q^{\ast}} =$ 300, 500, 700, and 900 GeV/$c^2$.}
\end{figure}

 Multijet background was simulated using the next-to-leading order
 (NLO) program {\sc jetrad}~\cite{jetrad}, with the {\sc
 cteq6m}~\cite{cteq6} PDFs, and renormalization scale ($\mu$) of
 0.5$E_T^{\rm max}$, where the $E_T^{\rm max}$ is the $E_T$ of the
 highest-$E_T$ parton.  Partons within $1.3 {\cal R}$ of one another
 are clustered into a single jet if they are within ${\cal R} = 0.7$
 of their $E_T$-weighted $\eta$,$\phi$ centroid~\cite{d0_dijet}.  The
 two highest-$E_T$ jets are used to calculate the dijet mass, which is
 then smeared using the measured mass resolutions. The resulting
 distribution is shown in Fig.~\ref{jetrad_fig}.

\begin{figure}[htb]
\includegraphics[width=\columnwidth ]{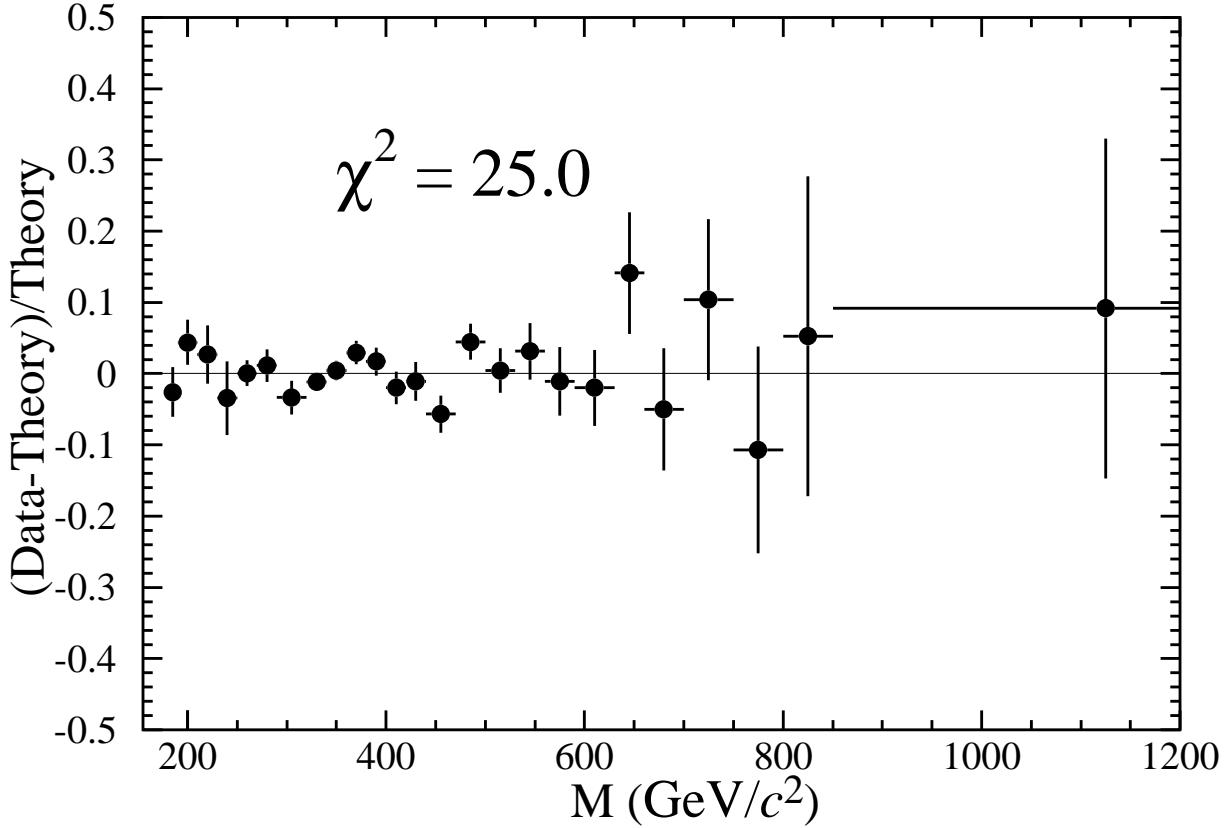}
\caption{\label{jetrad_fit_fig} The difference between data and the
 smeared {\sc Jetrad} NLO QCD prediction normalized to the theoretical
 prediction $((\mbox{Data}-\mbox{Theory})/\mbox{Theory})$ using the
 {\sc cteq6m} PDFs and a single renormalization scale $\mu = 0.5
 E_T^{\rm max}$. The vertical error bars represent the sum of the
 uncorrelated uncertainties added in quadrature, while the horizontal
 error bands represent the widths of the mass bins. The highest mass
 bin extends to 1400~\gevcc .}
\end{figure}

 A comparison between the background prediction and the data is given
 in Fig.~\ref{jetrad_fit_fig} (only uncorrelated uncertainties are
 shown). The $\chi^2$ of the comparison is 25.0 for 25 degrees of
 freedom.  This fit shows no obvious evidence for the existence of new
 particles.

\begin{figure}[htbp]
\includegraphics[width=10cm ]{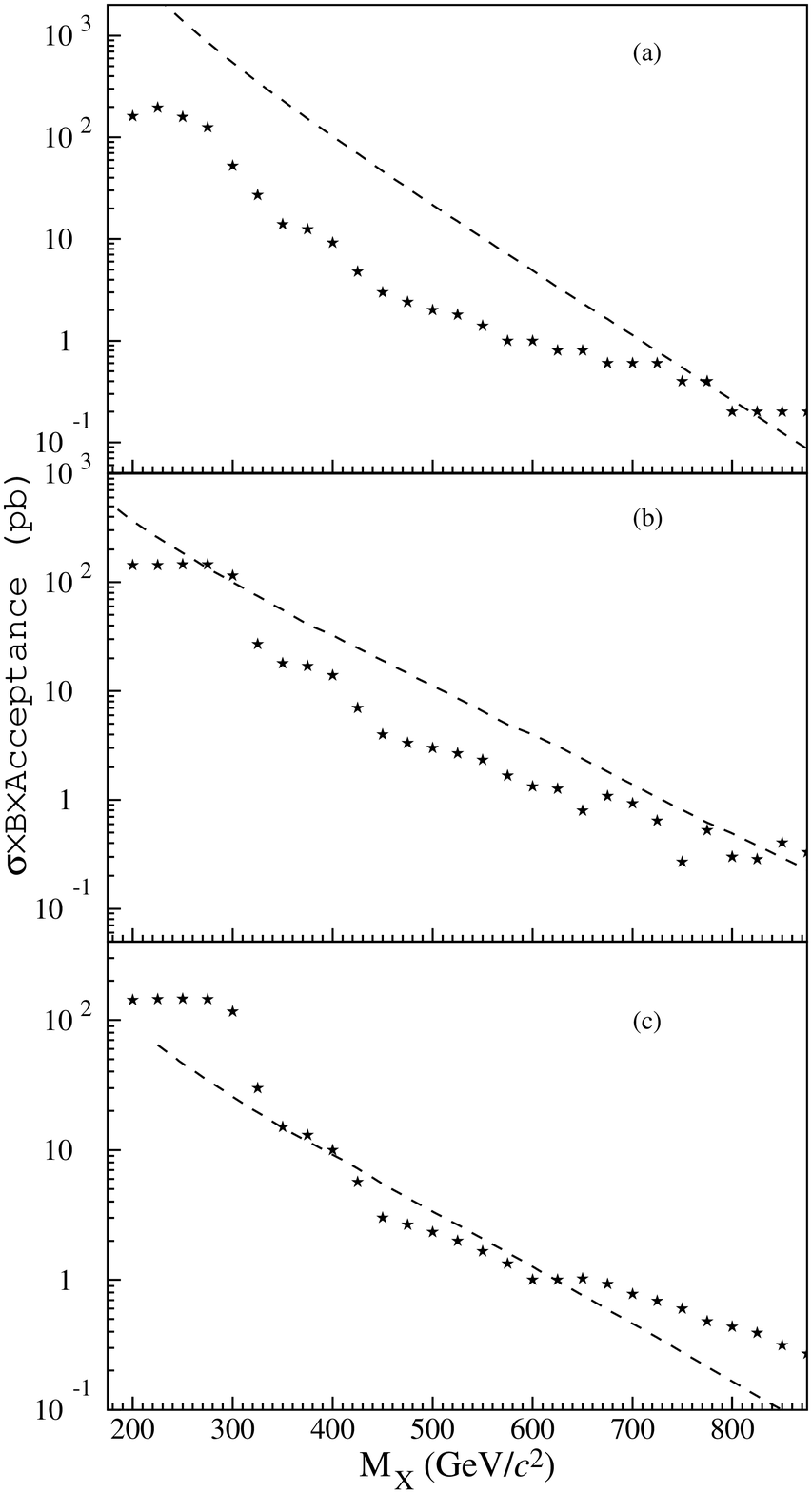}
\caption{\label{95_limit} The 95\% CL on the production cross section
  multiplied by $B(X \rightarrow \mbox{dijet})$ and acceptance, using
  the {\sc cteq6M} PDFs for: (a) an excited quark \qstar\ (stars),
  compared with the predicted cross section (dashed line);
  $M_{q^{\ast}} < 775$ GeV/$c^2$ is excluded; (b) similarly, for a
  \wprime\ boson (stars), $300 < M_{W^{\prime}} < 800$ GeV/$c^2$ is
  excluded; and (c) for a \zprime\ boson (stars), $400 <
  M_{Z^{\prime}} < 640$ GeV/$c^2$ is excluded.  }
\end{figure}

 The 95\% confidence level (CL) limits on the production cross
 sections for the three resonances are extracted using the same
 Bayesian method described above.  In Fig.~\ref{95_limit} we compare
 our measured 95\% CL limits (stars) with the expected cross section
 multiplied by the branching fraction ($B$) and acceptance for
 particles decaying to dijets (dashed curve). Branching fractions to
 all possible quark and gluon states are included in the acceptance.
 The acceptances for excited quarks (\wprime\ and \zprime\ bosons)
 range from 20\% at 200 \gevcc\ to 60\% (50\%) for masses above 700
 \gevcc . We exclude excited quarks with $M_{q^{\ast}} < 775$
 GeV/$c^2$. This is is the most restrictive limit on excited quark
 production to date. A \wprime\ boson is ruled out in the mass range
 $300 < M_{W^{\prime}} < 800$ GeV/$c^2$. Previous
 measurements~\cite{cdfwprime, d0wprime} have excluded a \wprime\
 boson with mass below 300~GeV/$c^2$; our new measurement therefore
 sets a far more stringent lower limit on a \wprime\ boson mass of
 800~\gevcc . A \zprime\ boson with mass between 400 and 640 \gevcc\
 is also excluded.

%
We thank the staffs at Fermilab and collaborating institutions, and
acknowledge support from the Department of Energy and National Science
Foundation (USA), Commissariat \` a L'Energie Atomique and
CNRS/Institut National de Physique Nucl\'eaire et de Physique des
Particules (France), Ministry for Science and Technology and Ministry
for Atomic Energy (Russia), CAPES, CNPq and FAPERJ (Brazil),
Departments of Atomic Energy and Science and Education (India),
Colciencias (Colombia), CONACyT (Mexico), Ministry of Education and
KOSEF (Korea), CONICET and UBACyT (Argentina), The Foundation for
Fundamental Research on Matter (The Netherlands), PPARC (United
Kingdom), Ministry of Education (Czech Republic), A.P.~Sloan
Foundation, and the Research Corporation.


\begin{thebibliography}{99}

%
\bibitem[*]{lehner}
Visitor from University of Zurich, Zurich, Switzerland.
\bibitem[\dag]{przybycien}
Visitor from Institute of Nuclear Physics, Krakow, Poland.
%
\vskip 0.25cm


\bibitem{excited_quarks}
  U.~Baur, I.~Hinchcliffe and D.~Zeppenfeld, Int.\ J.\ Mod.\ Phys.\
  A {\bf 2}, 1285 (1987);
  U.~Baur, M.~Spira and P.M. Zerwas, Phys. Rev. D {\bf 42}, 815 (1990).

\bibitem{ext_gauge}
  P.~Frampton and S.~Glashow, Phys. Lett B {\bf 190}, 157 (1987); 

\bibitem{UA2wjj}
  UA2 Collaboration, J.~Alitti {\em et al.}, Z. Phys. C {\bf 49}, 17 (1991).

\bibitem{ua2xjj}
  UA2 Collaboration, J.~Alitti {\em et al.}, Nucl. Phys. B {\bf 400}, 3 (1993).

\bibitem{cdfxjj}
  CDF Collaboration, F.~Abe {\em et al.}, Phys. Rev. D {\bf 55}, 5263 (1997).

\bibitem{d0_dijet}
  D\O\ Collaboration, B. Abbott {\em et al.}, Phys. Rev. D {\bf 64}, 032003 (2001).

\bibitem{d0_dijet2}
  D\O\ Collaboration, B. Abbott {\em et al.}, Phys. Rev. Lett. {\bf 82}, 2457 (1999). 

\bibitem{d0}
  D\O\ Collaboration, S.~Abachi {\em et al.},  Nucl. Instr. Methods. Phys. Res. A {\bf 338}, 185 (1994).

\bibitem{energy_scale}
  D\O\ Collaboration, B. Abbott {\em et al.}, Nucl. Instrum. Methods Phys. Res. A {\bf 424}, 352 (1999).


\bibitem{eppley} 
  R.~Hamberg, W.L.~van Neervan, and T.~Matsuura, Nucl. Phys. B {\bf 359}, 343 (1991)


\bibitem{pythia}
  T. Sjöstrand {\em et al.}, Computer Phys. Commun. {\bf 135}, 238  (2001).

\bibitem{cteq6} 
  J. Pumplin, {\em et al.},  J. High Energy Phys.  {\bf 0207} 12 (2002). 


\bibitem{d0conlimits} 
  I.~Bertram {\em et al.},  ``A Recipe for the Construction of Confidence Limits,''  
  FERMILAB-TM-210 (2000). 

\bibitem{jetrad} 
  W.T.~Giele, E.W.N. Glover, and D.A. Kosower,
  Phys. Rev. Lett. {\bf 73}, 2019 (1994).
 
\bibitem{cdfwprime}
  CDF Collaboration, T.~Affolder {\em et al.}, Phys. Rev. Lett. {\bf 87},  231803 (2001).

\bibitem{d0wprime}
  D\O\ Collaboration, S.~Abachi {\em et al.}, Phys. Rev. Lett. {\bf 76}, 3271 (1996).


\end{thebibliography}
\end{document}